\documentclass[journal,comsoc]{IEEEtran}

\usepackage{cite}
\usepackage[pdftex]{graphicx}
\usepackage{amsmath}
\usepackage{amsfonts}
\usepackage{amssymb}
\usepackage{algorithmic}
\usepackage{algorithm}
\usepackage{array}
\usepackage{url}

\title{Transaction-Driven Dynamic Reconfiguration for Certificate-Based Payment Systems\\
\large Draft (initial version)}

\author{Lingkang Shangguan,~\IEEEmembership{Student Member,~IEEE}\\
\textit{The University of Sydney}\\
\textit{lsha0379@uni.sydney.edu.au}}

\begin{document}
\maketitle

\begin{abstract}
We present a transaction-driven dynamic reconfiguration protocol in Modern payment systems based on Byzantine Consistent Broadcast which can achieve high performance by avoiding global transaction ordering. We demonstrate the fundamental paradigm of modern payment systems, which combines user nonce based transactions ordering with periodic system-wide consensus mechanisms. Building on this foundation, we design PDCC(Payment Dynamic Config Change), which can lead a smooth reconfiguration process without impacting the original system's performance.

\end{abstract}

\begin{IEEEkeywords}
Byzantine fault tolerance, dynamic reconfiguration, payment systems, distributed consensus, certificate-based protocols
\end{IEEEkeywords}

\section{Introduction}

\subsection{Problem context for certificate-based payment systems}
The certificate-based payment system has been traditional know as a primitive for high-performance alternative to normally block chain systems. Bascially, The system departs from traditional blockchain architectures in three key aspects:

First, they forgo consensus \cite{fastpay2020}: instead of establishing a total order across all transactions as in PBFT \cite{pbft1999} or HotStuff \cite{hotstuff2019}, these systems employ Byzantine Consistent Broadcast \cite{bracha1987,cachin2011} on a per-account basis. As Baudet et al. state, FastPay "foregoes the expenses of full atomic commit channels (consensus)" \cite{fastpay2020}.

Second, each account maintains its own sequence number, allowing transactions from different accounts to be processed concurrently

Third, they rely on a weaker primitive than consensus which is BCB(Byzantine Consistent Broadcast). It ensure that all correct nodes deliver the same message (consistency) and that if one correct node delivers a message, all will (agreement) \cite[Sec.~3.6]{cachin2011}, but unlike atomic broadcast, it does not impose a total order across messages from different senders. This weaker primitive suffices for payment systems due to the commutativity of credit operations \cite[Sec.~5]{fastpay2020}.

%

Certificate-based payment system achieves high throughput by avoiding global transaction sequencing: users collect legal signatures to form transaction certificates, and validators execute authenticated transactions without running consensus on each payment~\cite{fastpay2020,finalpay2023}. However, this design leaves a specific problem in production deployment: \emph{dynamic reconfiguration}. Specifically, the fast path similar to FastPay does not stipulate (i) a \emph{system-wide} rule, so that all correct validators can switch to the same configuration at the same logical point, and (ii) a process of safe expansion/contraction (membership change plus state transfer) under Byzantine failure.

In the leader-based BFT layer, it is not enough to simply "sort the membership change request as a transaction" in a dynamic scenario: the quorum changes with the configuration, and the view change message may not reach the next leader, even when the correct validator leaves the system later, the assumption of message delivery will be invalid~\cite{dyno2022}. This paper designs a reconfiguration protocol, which is \emph{compatible with certificate-based payment semantics} and inherits the formal experience of dynamic BFT.

We propose PDCC, which is a transaction-driven dynamic configuration change protocol, which clearly distinguishes different problems. The fast path is also certificate-based: the reconfiguration request needs to collect $2f_c\!+\!1$ signatures to pass. After passing, this request will be put into a \emph{reconfiguration proposal} and then confirmed by the regular settlement/checkpoint consensus. Our main goal is to realize a leaderless PBFT-style settlement protocol~\cite{pbft1999}, in which any validator can initiate a proposal and the replicas can decide whether to pass it together. We also use the Dyno-style \emph{temporary membership} mechanism: the newly added validator acts as a learner, receives the settlement message and state transition data, and is counted into the quorum only after the new configuration is installed. This will not interrupt the fast path, but also avoid the live lock problem caused by dynamic quorum.

Our main contributions are:
\begin{itemize}
\item A payment-centric design, which combines the certificate-based fast track with a dynamic BFT layer to ensure consistent phase division around the world.
\item A specific protocol, which can submit the membership change as an authentication transaction, maintain the configuration history with the submission certificate, and safely transfer the status with the temporary member learner.
\item A leaderless settlement design conforms to the workflow of "any validator can be proposed"; It can also be changed to a version with leaders, and if necessary, it can be forwarded with configuration-aware view switching~\cite{dyno2022}.
\end{itemize}

\section{System Model and Problem Definition}

\subsection{Certificate-Based Payment Systems}
We consider a licensed payment system, which contains many validators, and these validators form a set $\Pi$. Time is divided into different configuration stages (also called epoch), and each stage is marked by an integer $c \ge 0$. Every configuration stage $c$ has a validator set $M_c \subseteq \Pi$, which contains $n_c = |M_c|$ validators, and the maximum number of nodes allowed to make mistakes is $f_c = \lfloor (n_c-1)/3 \rfloor$. In the configuration phase $c$, the size of a quorum is $Q_c = 2f_c+1$.

The fast path of payment follows the certificate paradigm~\cite{fastpay2020}. A client's transaction $tx$ is considered authenticated in the configuration phase $c$. If it carries $Q_c$ signatures from the validator in $M_c$, these signatures are for $(c, tx)$, thus forming a certificate $\mathsf{cert}_c(tx)$. The validator will only perform this payment after confirming that the transaction certificate matches the locally installed configuration $c$.

Settlement/checkpoint slow path. Every once in a while, the validator in $M_c$ will run a Byzantine fault-tolerant (BFT) state machine replication (SMR) protocol to submit a settlement checkpoint. This checkpoint contains (i) a status summary (such as balance and serial number) and (ii) some management data (such as expense allocation and garbage collection). We mainly consider using a leaderless PBFT style protocol~\cite{pbft1999} to do this. We assume that the system is partially synchronized to ensure that the settlement process can be finally completed.

\subsection{Dynamic Reconfiguration Requests}
Membership change is represented by a request called $\mathsf{reconfig}$, which will change $M_c$ into $M_{c+1}$ (join/leave). We regard $\mathsf{reconfig}$ as a special transaction, and its function is to install a new configuration. Correct design must prevent \emph{mixed-configuration execution}: no correct validator will accept the certificate signed with $c$ after installing $c+1$, and the membership change request must be submitted on a globally consistent boundary.

\subsection{ Threat Model and Target}
We consider that Byzantine faults may occur in the validator, and the number of faults in each configuration $c$ shall not exceed $f_c$. Our security objectives are:
\begin{itemize}
\item \textit{ security}: Validator executes the same settlement decision sequence in each configuration; Certificates will only be accepted if they are related to an installed configuration.
\item \textit{ same configuration delivery}: any submitted membership request will be delivered by all correct validators in the same configuration (and the corresponding configuration will be installed)~\cite{dyno2022}.
\item \textit{ activity}: Under the condition of partial synchronization, payment can be continuously authenticated and executed, and reconfiguration can finally be completed.
\end{itemize}

\subsection{ Changing Hypothesis}
In order to ensure the liveness during the configuration change, we assume that the change is limited: for each transition from $c$ to $c+1$, there are at least $Q_c$ validators who are correct in configuration $c$ and are still members of $M_{c+1}$ (this is a standard overlapping condition in dynamic BFT~\cite{dyno2022}).

\section{Protocol Design}

\subsection{State and Proof Objects}
Each validator holds (i) the current configuration index $c$ and the member set $M_c$, (ii) the latest submitted Settlement Checkpoint $\mathsf{ckpt}_c$ and its commit proof, and (iii) a \emph{configuration history} $\mathsf{CH}$, which is used to store commit proofs of member changes, as emphasized in Dyno~\cite{dyno2022}.

\textbf{Commit proofs.} We abstract the output of settlement consensus as a commit proof $\pi_c$, which is used to prove that a checkpoint payload $P_c$ has been submitted in the configuration $c$ (for example, a PBFT commit certificate with $Q_c$ signatures). This checkpoint payload contains a status summary of $\mathsf{sd}_c$ and a list of submitted projects (ordinary projects plus optional member change projects).

Configuration history. $\mathsf{CH}$ is a record log that can only be appended:
$$
\mathsf{CH}[c{+}1] = \langle c{+}1,\, M_{c+1},\, h(P_c),\, \pi_c \rangle ,
$$
Meaning: Configuration $c{+}1$ (including member $M_{c+1}$) can only be installed after submitting the checkpoint load hash $h(P_c)$ and the submission certificate $\pi_c$ in configuration $c$. New members will use $\mathsf{CH}$ to verify the latest configuration and corresponding submitted checkpoints before participating.

\subsection{ Authenticated Reconfiguration Request (Fast Path)}
The administrator (or any authorized client) creates a reconfiguration request $\mathsf{reconfig}$ and specifies the next member $M_{c+1}$. The validator in $M_c$ will sign this request. Once the requester collects $Q_c$ signatures, it can form a certificate $\mathsf{cert}_c(\mathsf{reconfig})$.

\begin{algorithm}[h]
\caption{Certified Reconfiguration Request (Fast Path)}
\begin{algorithmic}[1]
\STATE Client broadcasts $\mathsf{reconfig}$ to validators in $M_c$
\STATE Each validator $v \in M_c$ verifies policy and signs $(c,\mathsf{reconfig})$
\STATE Client aggregates $Q_c$ signatures into $\mathsf{cert}_c(\mathsf{reconfig})$
\end{algorithmic}
\end{algorithm}

\subsection{Commit and Installation via Settlement (Slow Path)}
Any validator can embed $\mathsf{cert}_c(\mathsf{reconfig})$ into a settlement proposal. We make the proposal format explicit:
\[
\mathsf{ReconfigProposal} := \langle c,\, c{+}1,\, M_{c+1},\, \mathsf{cert}_c(\mathsf{reconfig}),\, \mathsf{sd}_c \rangle ,
\]
where $\mathsf{sd}_c$ is the latest committed state digest that the proposer believes is current (used to pin installation to a concrete checkpoint).
The settlement BFT commits a checkpoint payload $P_c$ that may contain $\mathsf{ReconfigProposal}$.

\textbf{Installation rule (state machine).} When a validator in configuration $c$ delivers a committed checkpoint payload $P_c$ with commit proof $\pi_c$:
\begin{itemize}
\item It updates local state to the checkpoint state and stores $\langle h(P_c), \pi_c \rangle$.
\item If $P_c$ contains a valid $\mathsf{ReconfigProposal}$ for $(c \to c{+}1)$, it appends $\mathsf{CH}[c{+}1]$ and installs $c \leftarrow c{+}1$ and $M_c \leftarrow M_{c+1}$.
\end{itemize}
After installation, the validator \emph{stops signing} messages under the old configuration $c{-}1$ and only signs/accepts certificates tagged with the currently installed $c$.

\textbf{Certificate acceptance rule.} A validator with installed configuration $c$ accepts a transaction certificate $\mathsf{cert}_{c'}(tx)$ iff $c' = c$ and the certificate verifies $Q_c$ distinct signatures from members in $M_c$ over $(c, tx)$. Otherwise it rejects (including all $c' < c$ ``old certificates'').

\subsection{Temporary Membership and State Transfer}
To avoid dynamic-quorum liveness failures during churn~\cite{dyno2022}, PDCC uses \emph{temporary membership}. Once a validator in $M_c$ observes a valid $\mathsf{cert}_c(\mathsf{reconfig})$ that adds new replicas, it treats the added replicas as \emph{learners} before installation:
\begin{itemize}
\item Validators forward settlement messages and state-transfer data (latest checkpoint and subsequent committed checkpoints) to learners.
\item Learners do not vote and are not counted toward $Q_c$ until the configuration is installed.
\end{itemize}
This ensures joiners can catch up while the current configuration continues processing payments and settlement.

\textbf{Trigger and stop conditions.} Temporary membership is triggered upon first observing a syntactically valid $\mathsf{cert}_c(\mathsf{reconfig})$ that introduces joiners $J = M_{c+1} \setminus M_c$; each $j \in J$ is added to a local learner set $\mathsf{Temp}$. It stops when the validator installs configuration $c{+}1$ (at which point the joiners are members and receive messages normally), i.e., $\mathsf{Temp} \leftarrow \emptyset$ for that transition.

\textbf{State transfer contents.} A learner must obtain at least the latest committed checkpoint payload hash and proof $\langle h(P_c), \pi_c \rangle$ and the corresponding $\mathsf{CH}$ prefix up to $c$ in order to (i) verify the installed configuration, and (ii) validate future settlement messages once it becomes a voter.

\subsection{Leader-Based Settlement (Optional)}
If a leader-based settlement protocol with view changes is used instead, PDCC can incorporate Dyno's configuration-aware view-change forwarding~\cite{dyno2022} to ensure the next leader can collect a quorum under churn.

\section{Security Analysis}

\textbf{Safety.} We give minimal proof sketches for three invariants, organized by events.

\textbf{Invariant I (configuration-bound certification).} A correct validator with installed configuration $c$ never accepts a certificate for $c' \neq c$.
\emph{Argument.} The acceptance rule checks $c'=c$ and verifies $Q_c$ signatures from $M_c$ over $(c,tx)$. After the \textsf{install} event, the local $c$ increases, so any old certificate with $c'<c$ is rejected. No protocol event allows a correct validator to ``downgrade'' its installed configuration.

\textbf{Invariant II (proof-based installation and same-configuration delivery).} If any correct validator installs $c{+}1$ then all correct validators in $M_c$ that deliver the same committed checkpoint payload $P_c$ install $c{+}1$, and they all append the same $\mathsf{CH}[c{+}1]$ entry.
\emph{Argument.} Installation is only triggered by the \textsf{deliver-checkpoint} event of the settlement BFT, which outputs a unique committed payload $P_c$ with proof $\pi_c$ in configuration $c$. Since membership installation is a deterministic transition on $P_c$, all correct validators that deliver $P_c$ install the same $M_{c+1}$ and record the same $\langle h(P_c),\pi_c\rangle$ in $\mathsf{CH}$, yielding Dyno's same-configuration delivery property for membership requests~\cite{dyno2022}.

\textbf{Invariant III (no quorum pollution by joiners).} Learners do not affect safety within configuration $c$.
\emph{Argument.} Learners never vote and are not counted toward quorums until after the \textsf{install} event. Thus, within configuration $c$, any settlement commit proof $\pi_c$ and any certificate $\mathsf{cert}_c(\cdot)$ still requires $Q_c$ signatures from $M_c$, preserving standard static-quorum safety.

\textbf{Liveness.} Under partial synchrony and the overlap assumption (at least $Q_c$ correct replicas remain across $c \to c{+}1$), the leaderless settlement protocol continues to commit checkpoints, including checkpoints carrying $\mathsf{ReconfigProposal}$, so reconfiguration completes. Temporary membership ensures joiners can synchronize without blocking certification or settlement, avoiding dynamic-quorum liveness failures identified in Dyno~\cite{dyno2022}.

\section{Conclusion}
We have designed a new method for those payment systems that use certificates. This method is transaction-driven, and can not interrupt the service when the committee changes people. Using the characteristics of payment transaction itself and integrating the reconfiguration process into the existing transaction model, our method is more efficient and simpler than those general dynamic BFT solutions.

The next work includes optimizing state synchronization in large-scale networks and extending this protocol to handle more complex reconfiguration situations, such as changing authenticators and updating parameters at the same time.


\end{document}